\documentclass[]{spiejour}
\usepackage[]{graphicx}
\newcommand{\farcs}{\mbox{\ensuremath{.\!\!^{\prime\prime}}}}
\title{NICI: combining coronagraphy, ADI, and SDI}

\author{\'Etienne Artigau,\supscr{a} Beth A. Biller,\supscr{b}  Zahed Wahhaj,\supscr{b} Markus Hartung,\supscr{a} Thomas L. Hayward,\supscr{a} Laird M. Close,\supscr{c} Mark R. Chun,\supscr{d} Michael C. Liu,\supscr{b} Gelys Trancho,\supscr{a} Fran\c cois Rigaut,\supscr{a} Douglas W. Toomey\supscr{e} and Christ Ftaclas\supscr{b}}

\affiliation{\supscr{a}Gemini Observatory, Southern Operations Center, Association of Universities for Research in Astronomy, Inc., Casilla 603, La Serena, Chile\\
  \supscr{b}Institute for Astronomy, University of Hawaii, 2680 Woodlawn Drive, Honolulu, Hawaii 96822, USA\\
  \supscr{c}Steward Observatory, University of Arizona, 933 North Cherry Avenue, Tucson, Arizona 85721, USA\\
  \supscr{d}Institute for Astronomy, University of Hawaii, 640 North A'ohoku Place, 209 Hilo, Hawaii 96720-2700, USA\\
  \supscr{e}Mauna Kea Infrared, LLC, 21 Pookela St., Hilo, Hawaii 96720, USA\\
}
\authorinfo{Send correspondence to \'Etienne Artigau: \linkable{eartigau@gemini.edu}}

\begin{document} 
  \maketitle 

\begin{abstract}
The Near-Infrared Coronagraphic Imager (NICI) is a high-contrast AO imager at the Gemini South telescope. The camera includes a coronagraphic mask and dual channel imaging for Spectral Differential Imaging (SDI). The instrument can also be used in a fixed Cassegrain Rotator mode for Angular Differential Imaging (ADI). While coronagraphy, SDI, and ADI have been applied  before in direct imaging searches for exoplanets. NICI represents the first time that these 3 techniques can be combined. We present preliminary NICI commissioning data using these techniques and show that combining SDI and ADI results in significant gains.
\end{abstract}


\section{INTRODUCTION}
\label{intro} 
The Near-Infrared Coronagraphic Imager (NICI) is a new instrument at Gemini-South Telescope\cite{Ftaclas2003}. NICI had its first light in February 2007 and is currently undergoing on-sky tests before embarking on an ambitious 50-night planet-search campaign. The campaign will be executed in queue mode and should significantly improve the statistical constraints on the presence of giant planets around nearby main-sequence stars. The campaign builds on similar surveys by Refs.~\cite{Biller2007, Lafreniere2007gdps, Lowrance2005}.

NICI combines a curvature AO system and a dual-channel imager. The two optical trains are designed to operate in a Spectral Differential Imaging (SDI) mode. In this mode the field is imaged in two close wavelengths selected such that any cold companion would have widely different flux in the two channels while the stellar spectral energy distribution remains relatively flat. This allows for a calibration of the stellar speckle pattern using the {\it off} channel with minimal self-subtraction of the companion. All previous on-sky applications of this mode \cite{Marois2005, Biller2007} as well as the nominal NICI mode use the deep methane absorption band around $1.6~\mu$m, but the technique could conceivably be used on other spectroscopic features. SDI is limitations by the decorrelation of speckles between the channels due to non-common path aberrations, differences in the filter bandpasses, wavelength difference between filters, etc. Once other noise contributions (readout noise, uncorrected atmospheric speckles, photon noise...) have averaged-out, no further gain in sensitivity can be obtained without resorting to other techniques as further integrations only provide higher signal-to-noise ration (S/N) on the fixed speckle pattern and doesn't help detect a planet buried within it.

The second technique envisioned for the NICI campaign and general high-contrast imaging is roll subtraction, a.k.a. Angular Differential Imaging (ADI) \cite{Liu2004, Marois2006}. The technique is similar to the roll subtraction used on HST \cite{Schneider2003} and takes advantage of the field rotation provided by an altitude-azimuth mount such as that of the Gemini South telescope. In this mode the Cassegrain focus or rotator is fixed, so the field rotates through the observation leaving all the optics (telescope, AO system, dual-imager) fixed. In this mode, field rotation provides a self-calibration of the PSF. For every image, an estimate of the PSF can be built using the rest of the dataset and subtracted from the frame of interest. Once this is done for all frames, they can be rotated to a common position angle (PA) before combination. As pointed out by Ref.~\cite{Marois2006}, any remaining speckle will add-up incoherently in the final image thus ensuring an ever increasing sensitivity.  As the field also rotates during integrations, there is a greater amount of image elongation at larger radii, and this field position-dependent loss in sensitivity needs to be taken into account when designing a survey strategy (see Biller et al. in this Volume). As a general rule, in the relatively short integrations used with NICI (typically a few seconds to a minute), the angular smear is small, especially in the scientifically most relevant part of the field (the inner $\sim2\farcs0$). ADI is most efficient in the outer parts of the PSF where relatively fast field rotation allows one to use images taken within minutes of the frame of interest to built the reference PSF. In the inner part of the field (less than $\sim1\farcs0$), the slower field rotation implies that only images taken tens of minutes from the frame of interest can be used to reconstruct the PSF estimate. The ever-evolving instrumental speckle pattern therefore implies that the ADI-reconstructed PSF will be better correlated at large separation than closer in. In some particularly unfavorable cases (i.e. objects low on the horizons or transiting directly overhead), ADI subtraction at small separations cannot be performed at all within a $1-2$ hours long observation. It is important to noted that ADI and SDI are {\it not} competing techniques and that they may very well be used on the same dataset.

The third important feature of NICI is the presence of a semi-transparent focal plane mask. A wheel allows the positioning of a set of masks ranging in diameter from $0\farcs22$ to $0\farcs90$ as well a clear glass element of the same optical thickness. The central part of masks have a typical transmission of 1 part in 300 (See Ftaclas et al. in this Volume). While the Strehl ratio obtained by NICI ($\sim0.3$ in $H$, $\sim0.5$ in $K$) are not sufficient to perform diffraction coronagraphy, the focal plane masks still help improve sensitivity by 1) limiting the amount of light entering the system and thus minimizing ghosts, 2) avoiding the deep saturation of the central parts of the PSF that would create electronic ghosts and leakage and 3) providing a non-saturated diffraction core as both an astrometric and photometric reference for registering images.

\begin{figure}
   \begin{center}
   \begin{tabular}{cc}
   \includegraphics[height=5.5cm]{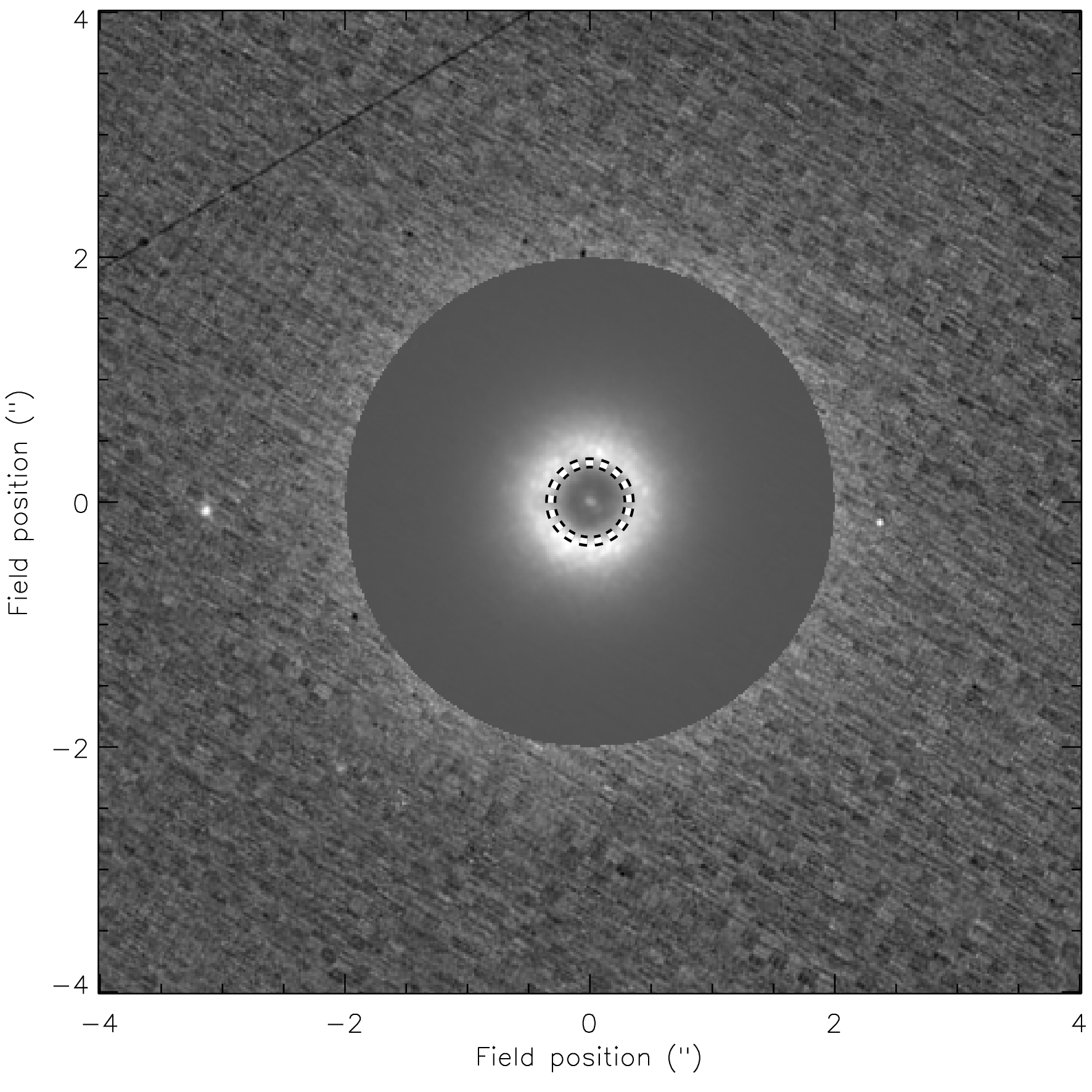} &
   \includegraphics[height=5.5cm]{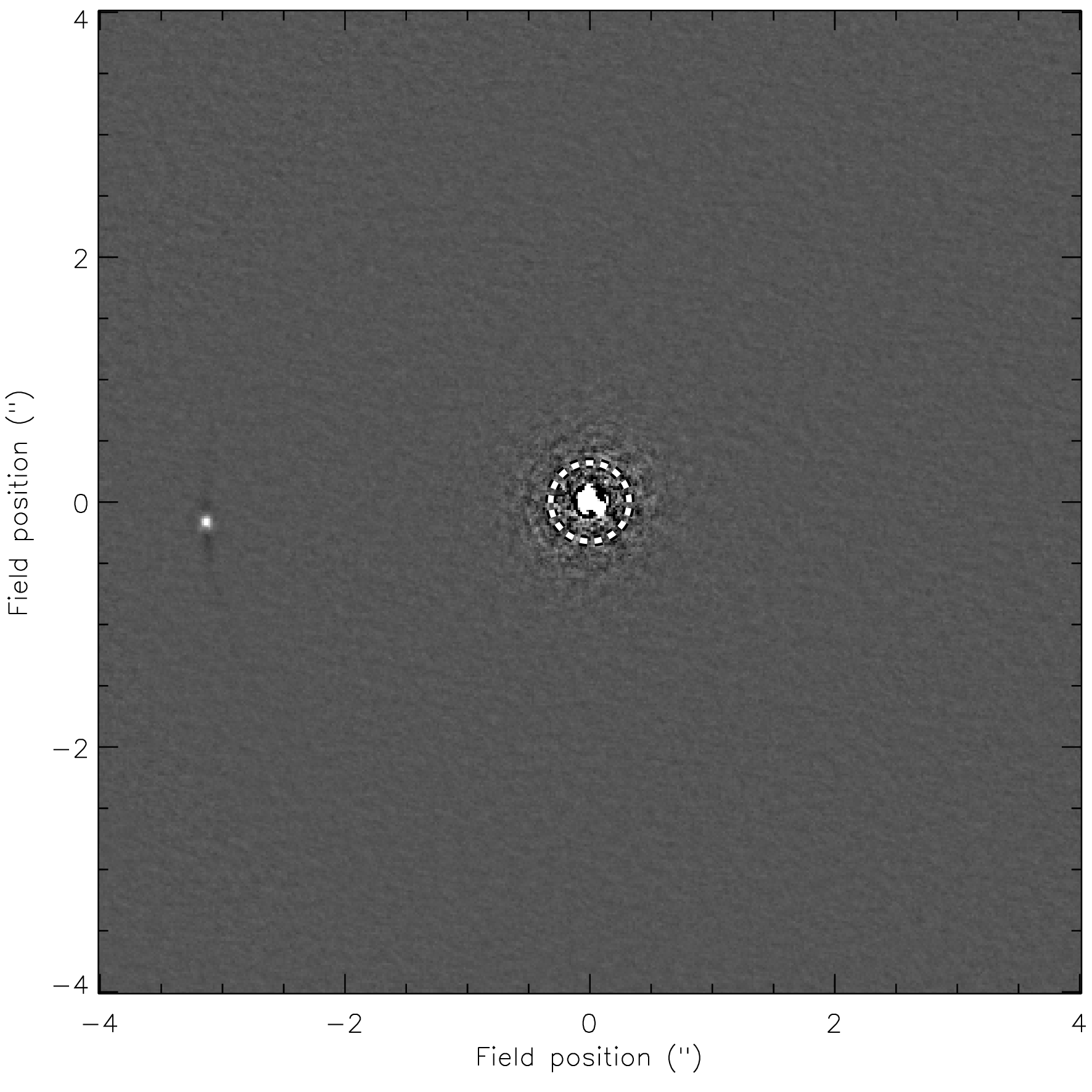} \\
   (a) & (b)
   \end{tabular}
   \end{center}
   \caption[example] 
   {\label{beforeafter} 
Left panel shows a single NICI image of TWA-7 after dark-subtraction and flat-fielding. The image inside a $2\farcs0$ radius has been divided by 30 to account for the large dynamic range and show speckles in the inner parts of the point spread function. The right panel shows the same field after the complete data reduction, for a total integration time of 31 minutes. For both panels, the scale ranges from $-0.2$ to $1\times$ the peak of the background object $\sim3\farcs1$ from TWA-7. This object is $\Delta H=9.4$ mag fainter than TWA-7. A dashed circle indicates the edge of the $0\farcs32$-wide occulting mask in both panels.
} 
   \end{figure}

 This paper describes the broad data reduction of a preliminary NICI dataset obtained during commissioning and provides sensitivity measurements for important steps to quantify the relative gains. Observations are described in \S~\ref{observations}, \S~\ref{reduction} describes the various data reduction steps and \S~\ref{discussion} details their relative sensitivity gains. Figure~\ref{beforeafter} shows an example of a raw NICI frame and the final image obtained with the data reduction method described here.

\section{Observations}\label{observations} 
 The dataset used here was taken on TWA-7 ($V=11.1$, $H=7.1$), a young star with a nearby ($\sim3\farcs1$ at the time of observation, $2\farcs445$ in the 1998 HST discovery images \cite{Neuhauser2000}) faint point source that was once a good planetary companion candidate but is now known to be a mere background object. The known background provides a welcome sanity check of various data-reduction steps. This dataset has been taken with the Cassegrain rotator fixed and serves as a comparison with Cassegrain rotator tracking dataset taken immediately afterward (see Biller et al., this Volume). The halo at the edge of the occulting mask is at a fifth of the dynamical range of the detector in a $30$ s exposure, still leaving some room for significantly longer integrations or brighter stars with a similar exposure time. A $0\farcs32$ occulting mask was used with a $90\%$ pupil mask and beam splitting was done using the $50\%/50\%$ beam splitter. The {\it short} and {\it long} wavelength channels respectively used filters centered at 1.578~$\mu$m and 1.652~$\mu$m with $4\%$ and $3.5\%$ fractional bandpasses. At a declination of $\delta=-33.7^{\circ}$, TWA-7 passes within a few degrees of zenith as seen from Cerro Pach\'{o}n (latitude of $-30.2^{\circ}$) providing very fast field rotation at transit. Our 62 30-s integration frames span 1.1 hours for a total field rotation of $74.5^{\circ}$. Such a field rotation is enough to build the reference PSFs needed for the ADI subtraction at all separations beyond the edge of the occulting mask.

\section{Data reduction}
\label{reduction}
\subsection{Preliminary steps}
\label{cosmetic}
The first data reduction steps, common to most astronomical imaging, involve a dark subtraction and flat fielding. The flat field is obtained using the Gemini calibration module (GCAL, \cite{RamsayHowat2000}) using a flat illumination and is built from the median combination of the difference between illuminated and non-illuminated frames. The NICI pupil stop and very narrow field of view ensures that the detector illumination is the same as for on-sky observations. Dark frames are obtained using blocking filters in both wheels. Sky frames are not used as 1) all non electronic (i.e. not seen in dark frames) contributions are low spatial frequency and will be removed by high-pass filtering and 2) the ADI subtraction provides a self-consistent sky subtraction. NICI detectors have a slight but significant relative rotation of $1.15^{\circ}$ and scale difference of $\sim0.4\%$ as determined by images of a focal plane pinhole grid; images are rotated, scaled and registered. Accurate registering of the images requires special care, and the most accurate results were found by first registering on the diffraction core and fine-tuning the measurement using a cross-correlation of the speckle patterns between the two channels.

\subsection{High-pass filtering}
\begin{figure}
   \begin{center}
   \begin{tabular}{c}
   \includegraphics[height=10.5cm]{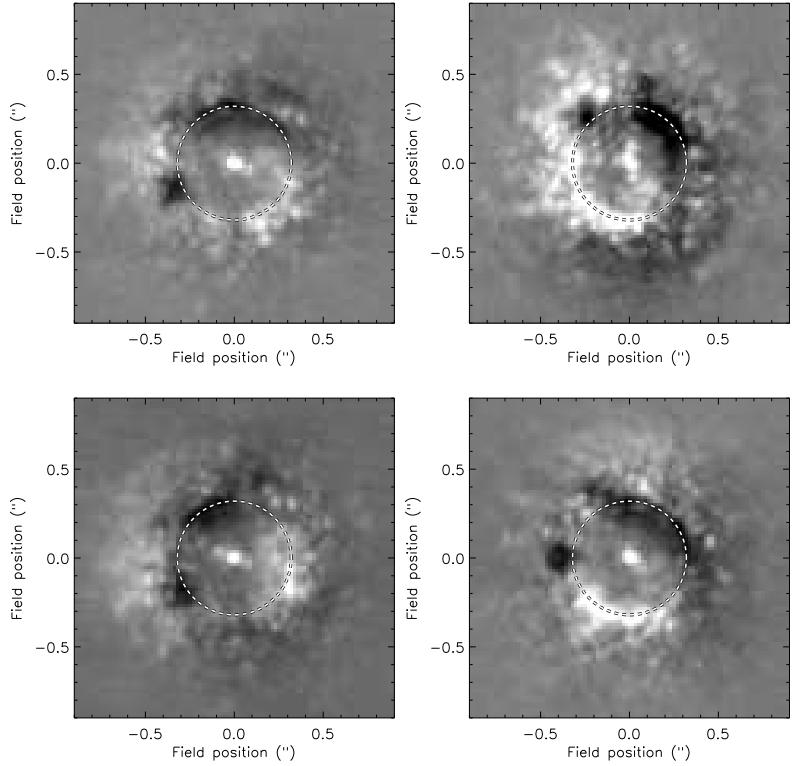}
   \end{tabular}
   \end{center}
   \caption[example] 
   { \label{halo} 
Four representative short-channel images. Images have been dark-subtracted, flat-fielded and a median radial profile has been subtracted. Structure variations are seen on various physical scales, both in the low spatial frequency of the halo and in the speckle structure. } 
   \end{figure}

The next step in the data reduction is a high-pass filtering of all images. This serves two purposes; first the spatial scale that contains the planets is known, so the images are filtered as to preserve only spatial scales similar to those of point sources. Secondly, the low-frequency component will vary from frame to frame; the Strehl ratio will vary with changing seeing and the halos of different amplitudes hamper the filtering of FWHM-size speckles. Figure~\ref{halo} shows four individual images after the steps described in the previous section and a radial profile subtraction. The high-pass filtering was done in two steps; first a median radial profile was subtracted from individual frames, then a high-pass filter was applied to the frame. The low-passed image subtracted was smoothed using a $11\times11$ median box. Median filtering is used because it better rejects bad pixels and cosmic rays, but a boxcar filtering gives basically identical sensitivity results.

\subsection{Angular Differential Imaging LOCI}
The Locally Optimized Combination of Images (LOCI) algorithm described by Ref.~\cite{Lafreniere2007loci} is used here to construct an estimate of the PSF for every image and minimize speckle noise that hampers faint companion detection. In this technique, for a given region of the image of interest (in Ref.~\cite{Lafreniere2007loci} as well as our implementation, annular or segment of annuli regions). For this region, a set of reference images is defined and linearly combined as to minimize the root-mean-square of the difference between the image of interest and the reconstructed image. Using the Ref.~\cite{Lafreniere2007loci} notation, this minimization is expressed as :
\begin{equation}
  \begin{array}{c}\sigma^2=\sum_{i}{ \left(O^{T}_{i} - \sum_{k}{c^{k}O^{k}_{i}}\right)^2  }\end{array}
\end{equation}

where $i$ is the pixel index in the region of interest, $O^{T}$ the region of interest in the image for which the reference PSF is reconstructed, $O^{k}$ is the same region for the $k^{\rm th}$ reference image and $c^{k}$ is the weight associated with the the $O^{k}$. Ultimately one wants to find the set of  $c^{k}$ values that will minimize $\sigma^2$. This minimization can be expressed as a simple linear problem, and the $c^{k}$ weight vector can be determined using :

\begin{equation}
  c=M^{-1}*v\, {\rm, where }
\end{equation}

\begin{equation}
  M=\left[\begin{array}{cccc}
      \sum_{i}{O^{k=1}_{i}O^{k=1}_{i}} & \sum_{i}{O^{k=1}_{i}O^{k=2}_{i}} & ... & \sum_{i}{O^{k=1}_{i}O^{k=N}_{i}}\\
      \sum_{i}{O^{k=2}_{i}O^{k=1}_{i}} & \sum_{i}{O^{k=1}_{i}O^{k=2}_{i}} & ... & \sum_{i}{O^{k=2}_{i}O^{k=N}_{i}}\\
      ... & ...& ... & ... \\
      \sum_{i}{O^{k=N}_{i}O^{k=1}_{i}} & \sum_{i}{O^{k=N}_{i}O^{k=2}_{i}} & ... & \sum_{i}{O^{k=N}_{i}O^{k=N}_{i}}
    \end{array}\right]\, {\rm and}
\end{equation}

\begin{equation}
  v=\left[\begin{array}{c}
      \sum_{i}{O^{T}_{i}O^{k=1}_{i}} \\
      \sum_{i}{O^{T}_{i}O^{k=1}_{i}} \\
      ...\\
      \sum_{i}{O^{T}_{i}O^{k=1}_{i}} 
    \end{array}\right]\, {\rm .}
\end{equation}

In the ADI case, all images for which the a source at the radii of the $O^{T}$ region has moved a source by more than 1.5 FWHM are used to built $O^{k}$. Including images taken with smaller field rotation would result in a self-subtraction of any companion. Having a significantly larger exclusion criteria would remove the images closest in time to $O^{T}$, which, depending on the instrumental speckle lifetime, are expected to show the highest degree of correlation with the image of interest. The number of images excluded from $O^{k}$ is a function of radius and increases at larger separations.

Once the LOCI-estimated PSFs have been subtracted for all frames in a given channel, the frames are rotated to a common position angle.

\subsection{Two-channel LOCI}
\label{tcloci}
As mentioned in Ref.~\cite{Lafreniere2007loci}, the LOCI algorithm can be generalized to any set of PSF that can serve as a linear basis for reconstructing a reference PSF of a given frame. In principle, the algorithm has been applied to images taken at different position angles (as in ADI) but could be used with images taken at different wavelengths, polarization states or even on reference stars. Here, the {\it short} channel dataset is processed in the same way as in the LOCI-subtraction described above but to the set of all images within the {\it short} channel with sufficient field rotation to avoid self-subtraction of companions, we add all images from the {\it long} channel. Any planetary companion will be cold enough to show deep CH$_4$ bands and will be much fainter in the {\it long} channel. We can therefore include all {\it long} channel images, including the one taken simultaneously with $O^{T}$. Before including all {\it long} channel images to the $O^{k}$ set, the images have to be scaled down by the ratio of the central wavelengths of the filter used to account for the wavelength dependency of the speckles. We note that this scaling alone ensures that beyond a field position of $\frac{\lambda_1}{D}\times\frac{\lambda_1}{\lambda_2-\lambda_1}\sim1^{\prime\prime}$ ($D$, diameter of the entrance pupil, $\lambda_1$, $\lambda_2$ are the {\it short} and {\it long} channel wavelengths) the scaling moves an object in the {\it long} channel by an amount sufficient to avoid self-subtraction, thus enabling the detection of non-methane bearing objects.

\section{Sensitivity \& Discussion}\label{discussion}
\begin{figure}
   \begin{center}
   \begin{tabular}{c}
   \includegraphics[height=10.5cm]{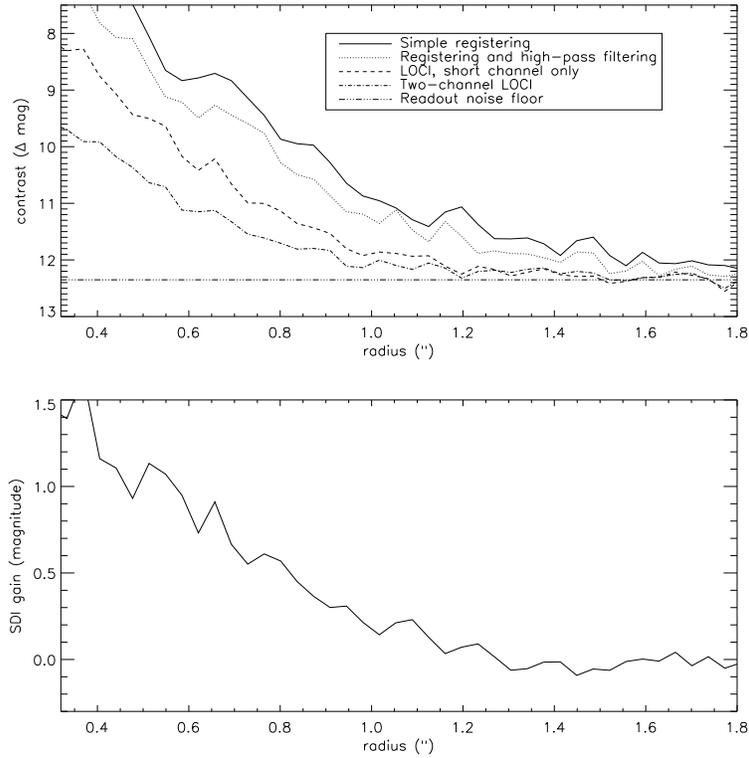}
   \end{tabular}
   \end{center}
   \caption[example] 
   { \label{contrast} 
Upper panel : The $5 \sigma$ contrast limits for the TWA-7 ($V=11.1$, $H=7.1$) dataset, totalling 31 minutes of integration, obtained after: a simple registering of flat-fielded, dark-subtracted images (solid curve); a combination of the radial-profile and high-pass subtracted images (dotted curve); a {\it short}-channel LOCI (dashed line); and the complete two-channel LOCI (dashed-dot line). The line at a $\Delta$~mag$=12.34$ ($H=19.4$) indicates the limit set by readout noise. This limit could be further lowered by $\sim$$1~$mag by using 16 or 32 NDRs instead of the 4 NDRs used in the dataset presented here. Lower panel~: difference between the  {\it short}-channel LOCI and two-channel LOCI curves. This difference gives a measure of the sensitivity obtained through spectral differential imaging.} 
   \end{figure}

The median-combined and PA-aligned images at various steps of the data reduction have been produced in order to quantify the relative gains of the various steps. Figure~\ref{contrast} illustrate both the resulting images and the $5 \sigma$ detection limits at the corresponding steps. Sensitivity curves shown in Figure~\ref{contrast} have been derived using a circular aperture with a 3.0~pixels radius and a sky annulus ranging from 6 to 12~pixels. Photometric calibration has been done using the background object mentioned earlier. The derived sensitivity curves is only weakly dependant on the aperture used, and other techniques such as convolving the images by a Gaussian with a 5~pixels FWHM and deriving a sensitivity curve using the RMS of the residual in annuli give similar results.

   \begin{figure}
   \begin{center}
   \begin{tabular}{c}
   \includegraphics[height=10.5cm]{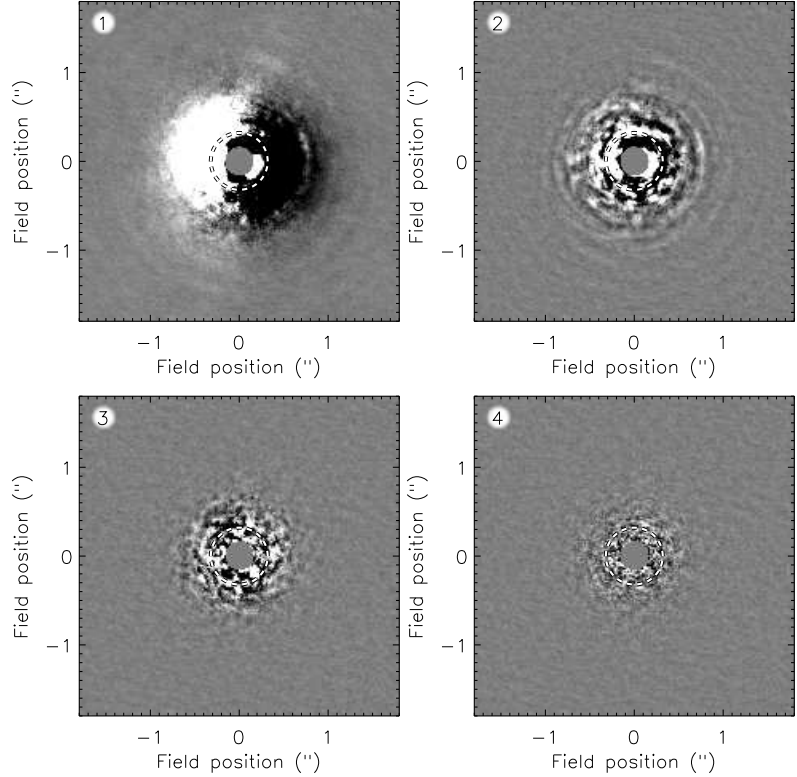}
   \end{tabular}
   \end{center}
   \caption[example] 
   { \label{combo} 
Median combination of the complete, PA-registered, dataset at various data reduction steps. For each image the corresponding sensitivity curve in Figure~\ref{contrast} is given. Image~\#1 shows the combined image after dark-subtract, flat-fielding and median radial-profile subtraction (solid line); this corresponds to the data reduction step illustrated by Figure~\ref{halo}. The strong asymmetry seen in the radial-profile subtracted image is readily seen in the image before subtraction and is not an artifact due to a misregistering of the radial profile (which would give a similar signature). Image~\#2 is the combination of the dataset after performing the high-passed filtering (dotted line). Image~\#3 shows the combined {\it short}-channel LOCI image (dashed line). Image~\#4 shows the final two-channel-LOCI image (dash-dot line). All images have a linear scaling ranging from $-0.5$ to $+0.5\times$ the peak value of the $\Delta H\sim9.4$ background object $3\farcs1$ from TWA-7.} 
   \end{figure} 

An important measure of the performance of NICI is the sensitivity gain through SDI. Implementing the dual channel imager has a financial, technical and sensitivity cost, part of the optical train has to be duplicated, in the design adopted here two detectors were used, $50\%$ of the light is lost at the beam splitter compared to a simple imager, splitting the light in two channels implies that the effective readout noise contribution is doubled relative to the signal, etc. Of course, this price is paid to obtained as to obtain better calibration of the speckle pattern and, at least for the most scientifically relevant part of the field, is providing a gain in sensitivity.

The difference between the {\it short}-channel LOCI and the dual-channel LOCI provides an estimate of the sensitivity gain in having two channels (see lower panel of Figure~\ref{contrast}). The final {\it short}-channel LOCI is the deepest image one could achieve using a single channel image. In a survey using two epochs to identify faint companions (such as the GDPS by Ref.~\cite{Lafreniere2007gdps}), this would define the sensitivity for a given epoch. One could argue that a single-channel imager would go $\sim0.7$~mag deeper within the same period of time thanks to the $50\%$ gain in throughput. This is certainly the case in the read-out noise limited part of the image beyond $\sim1\farcs2$, but the signal-to-noise in the inner part of the PSF is {\it not} limited by flux level, but by residual speckle noise that the LOCI algorithm failed to subtract. This can be seen in Figure~\ref{contrast}, the noise in the central part of the {\it short}-channel LOCI image is dominated by FWHM-sized structures, not pixel-to-pixel fluctuations. Gains in sensitivity for this region of the PSF can only be obtained by observing a larger number of individual realizations of the speckle pattern, {\it not} by increasing the per-image flux. At the edge of the occulting mask, there is a $\sim1.5$~mag sensitivity gain in including the second channel images, and this gain drops to about $0.2$~mag at a $1\farcs0$ radii. This rapid drop is easily explained by the fact that the efficiency of angular differential imaging increases with radii.



\acknowledgments 
The authors would like to thank the Gemini Engineering group for their hard work and dedication to the NICI project; Javier Luhrs, Ramon Galvez and Cristian Urrutia on software; Gaston Gausachs on NICI thermal enclosure.
\bibliographystyle{spiejour}


\end{document}